\documentclass[conference]{IEEEtran}
\IEEEoverridecommandlockouts
\usepackage{cite}
\usepackage{amsmath,amssymb,amsfonts}
\usepackage{algorithmic}
\usepackage{graphicx}
\usepackage{textcomp}
\usepackage{xcolor}

\usepackage{amsthm}
\usepackage{bbm}
\usepackage{mathrsfs}
\usepackage[linesnumbered,ruled,vlined]{algorithm2e}
\usepackage{dsfont}

\def\BibTeX{{\rm B\kern-.05em{\sc i\kern-.025em b}\kern-.08em
    T\kern-.1667em\lower.7ex\hbox{E}\kern-.125emX}}
\begin{document}

\title{Quality-Cost Trade-off on Constructing Logical Views for Vehicular Cyber-Physical Systems: A Deep Reinforcement Learning Approach\\
}

\author{\IEEEauthorblockN{Junyuan Wu$^1$, Xincao Xu$^2$, Chuzhao Li$^3$, Hao Zhang$^4$, Ke Xiao$^5$, Kai Liu$^{1, 3}$}
\IEEEauthorblockA{
$^1$College of Computer Science, Chongqing University, China \\
$^2$Shenzhen Institute for Advanced Study, University of Electronic Science and Technology of China, China\\
$^3$National Elite Institute of Engineering, Chongqing University, China \\
$^4$College of Computer Science and Technology, Chongqing University of Posts and Telecommunications, China\\
$^5$College of Computer and Information Science, Chongqing Normal University, China\\
Email: jy.wu@cqu.edu.cn, xc.xu@uestc.edu.cn, 
lichuzhao@126.com, 
zhanghao@cqupt.edu.cn, \\
xiaoke@cqnu.edu.cn,
liukai0807@cqu.edu.cn
}
\thanks{Xincao Xu is the corresponding author.}
}

\maketitle

\begin{abstract}
With the development of sensing technologies, vehicle-to-everything (V2X) communications, edge computing paradigm, vehicular cyber-physical systems (VCPS) are emerging as the most fundamental platform for realizing future intelligent transportation systems (ITSs).
In particular, the construction of logical views at the edge nodes based on heterogeneous information sensing and uploading are critical to the realization of VCPS.
However, a higher-quality view in terms of timeliness and accuracy may require higher cost on sensing and uploading.
In view of this, this paper is dedicated to striking a balance between the quality and the cost for constructing logical views of VCPS.
Specifically, we first derive an information sensing model based on multi-class M/G/1 priority queue and a data uploading model based on reliability-guaranteed vehicle-to-infrastructure (V2I) communications.
On this basis, we design two metrics, namely, age of view (AoV) and cost of view (CoV), simultaneously.
Then, we formulate a bi-objective problem to maximize the AoV and minimize the CoV.
Further, we propose a distributed distributional deep deterministic policy gradient (D4PG) solution to determine sensing information, frequency, uploading priority, transmission power, and V2I bandwidth.
Finally, we build a simulation model and give a comprehensive performance evaluation, and the simulation results conclusively demonstrate the superiority of the proposed solution.
\end{abstract}

\begin{IEEEkeywords}
Vehicular cyber-physical system, Cooperative sensing, Resource allocation, Deep reinforcement learning
\end{IEEEkeywords}

\section{Introduction}
\begin{figure}
\centering
  \includegraphics[width=0.99\columnwidth]{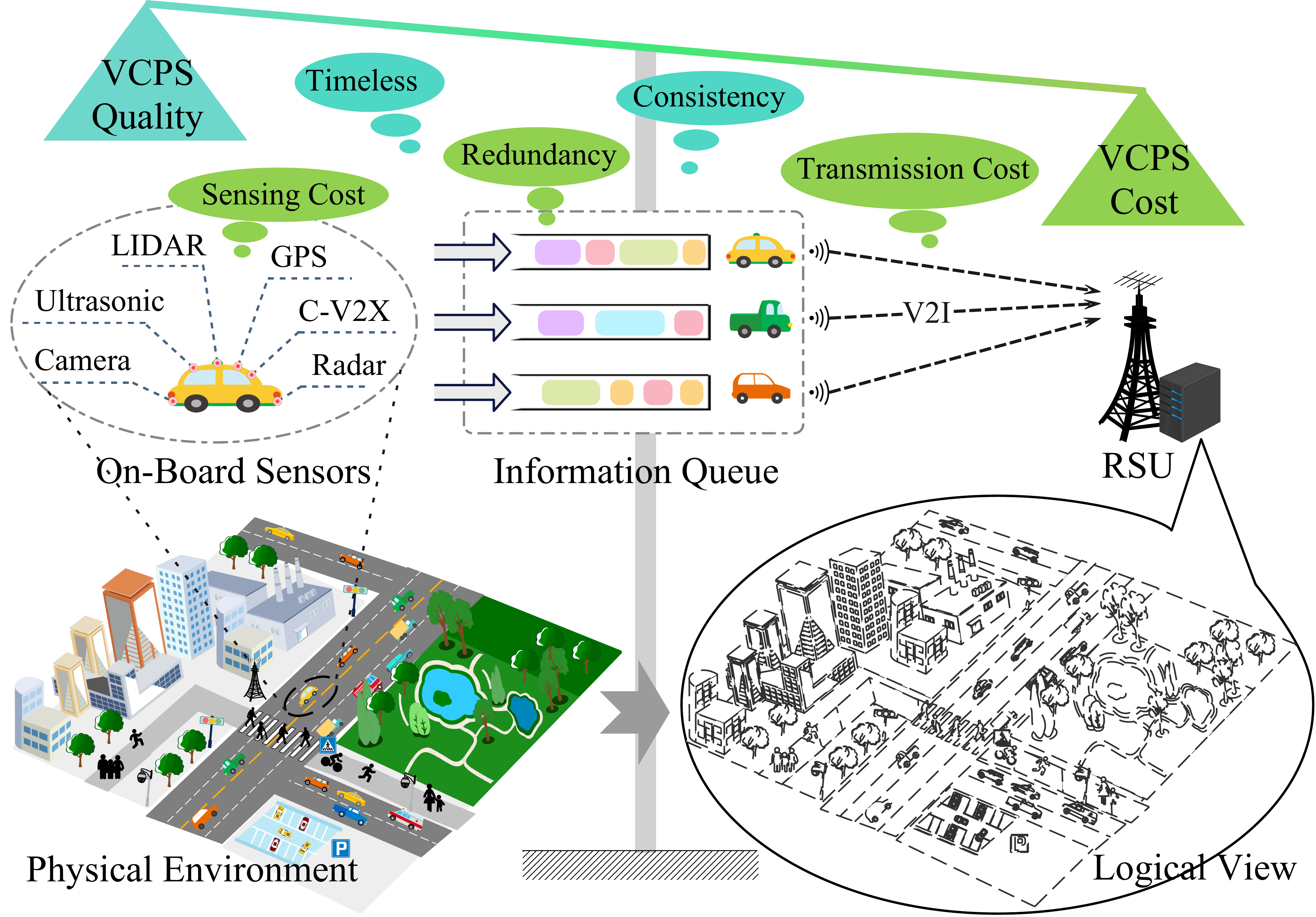}
  \caption{System model}
  \label{fig_0_system_example}
\end{figure}

Recent advances in sensing, communication, computing drive the development of vehicular cyber-physical systems (VCPS), which is a key enabler of the next generation of intelligent transportation systems (ITSs)\cite{xu2022age}.
As shown in Fig. \ref{fig_0_system_example}, vehicles may collaboratively sense via on-board sensors, such as GPS, cameras, radar, and LiDAR.
The heterogeneous information, including traffic light status, vehicle locations, point cloud data, and surveillance videos, are uploaded to the nearby roadside units (RSUs) by vehicle-to-infrastructure (V2I) communications.
Such information can be further used to construct logical views of the physical vehicular environment, which are critical for predicting, scheduling, and controlling in various upper-layer applications.
A high-quality view can reflect the physical vehicular environment accurately in real-time. 
However, a higher-quality view in terms of timeliness and accuracy may require higher cost on sensing and uploading, which increases additional overhead on energy consumption and information processing.
Therefore, it is crucial to construct high-quality and low-cost VCPS.

With continuously increasing attention on the quality and cost of VCPS, a trajectory-driven opportunistic routing (TDOR) protocol for message node delivery was proposed to reduce redundant routing overhead in VCPS   \cite{cao2018trajectory} and a fuzzy based channel selection in multichannel VCPS environments was proposed \cite{kasana2018fuzzy}.
On the other hand, great efforts have been devoted to data dissemination in vehicular networks, such as vehicular end-edge-cloud cooperative data dissemination architecture \cite{liu2020fog} and intent-based network control framework\cite{singh2020intent}.
To improve caching efficiency, some other researchers proposed content caching frameworks in vehicular networks, such as blockchain-empowered distributed content caching framework \cite{dai2020deep} and dynamic content caching scheme based on the cooperation among RSUs \cite{su2018edge}.
Some researchers studied task offloading mechanisms,  multi-dimensions intent-aware task offloading strategy \cite{liao2020learning} in vehicular networks and a joint task offloading and resource optimization via V2I communications method was proposed to maximize the service ratio\cite{xu2023joint}.
These studies on data dissemination, information caching, and task offloading formed the basis of modeling VCPS.
Further, several studies have been studied on predicting technologies, such as the hybrid velocity-profile prediction method \cite{zhang2019cyber}, lane-level localization and acceleration prediction \cite{zhang2020data}. 
Some researchers developed scheduling schemes, such as physical-ratio-K interference model-based broadcast scheme \cite{li2020cyber} and path planning scheduling method based on an established map model \cite{lian2020cyber}.
In addition, some studies proposed controlling algorithms, such as vehicle acceleration control algorithm \cite{lv2018driving} and collision warning system \cite{xu2020vehicular}.
These predicting, scheduling and controlling technologies facilitated the implementation of various upper-layer applications.
A few studies have concerned the information quality and transmission cost in VCPS, including timeliness \cite{liu2014temporal, dai2019temporal}, consistency \cite{liu2014realtime}, and accuracy \cite{yoon2021performance}. 
Nevertheless, to the best of our knowledge, none of the prior studies have investigated the trade-off between the quality and cost for constructing logical views of VCPS.

With above motivations, we present a scheduling algorithm  to striking a balance between the quality and the cost for constructing logical views of VCPS.
The primary contributions are summarized as follows. 
\begin{itemize}
\item We formulate the problem with the objectives of maximizing the quality and minimizing the cost of VCPS. Specifically, we derive an information sensing model and a data uploading model based on the multi-class M/G/1 priority queue and reliability-guaranteed V2I communications. Then,  two metrics named age of view (AoV) and cost of view (CoV) are defined to quantify the quality and cost of logical views, respectively. 
\item We propose a distributed distributional deep deterministic policy gradient (D4PG) solution. Specifically, the D4PG is implemented in the RSU with the action space of determining the sensing information, sensing frequencies, uploading priorities, transmission power, and V2I bandwidth allocation. The reward function is defined as the sum of the complement of achieved average AoV and average CoV in vehicular networks.
\item We give comprehensive performance evaluation. First, we build the simulation model based on real-world vehicular trajectories extracted from Didi GAIA Initiative \cite{didi}. We implement the proposed solution, and two competitive algorithms, including random allocation and multi-agent deep deterministic policy gradient (MADDPG) \cite{zhang2021adaptive}. The simulation results conclusively demonstrate the superiority of the proposed solution. In particular, D4PG outperforms RA and MADDPG by around 20.79\% and 13.88\%, respectively, in terms of maximizing the cumulative reward.
\end{itemize}

The remainder of this paper is organized as follows. 
Section II presents the system model.
Section III proposes the D4PG solution.
Section IV presents the numerical results. 
Finally, Section V concludes this paper.

\section{System Model}

\subsection{Information Sensing Model}
We consider the vehicular networks with $E$ RSUs and $S$ vehicles.
We denote the set of discrete time slots as $T$ and the set of heterogeneous information as $D$.
Each information $d \in D$ is characterized by a three-tuple $d=\left(\operatorname{type}_d, u_d, \left|d\right| \right)$, where $\operatorname{type}_d$, $u_d$, and $\left|d\right|$ are the type, state update interval, and size, respectively. 
Each RSU $e \in E$ is characterized by a three-tuple $e=\left (l_e, r_e, b_e \right)$, where $l_{e}$, $r_{e}$, and $b_{e}$ are the location, communication range, and bandwidth, respectively.
Each vehicle $s \in S$ is characterized by a three-tuple $s=\left (l_s^t, D_s, \pi_s \right )$, where $l_s^t$, $D_s$, and $\pi_s$ are the location, sensed information set, and transmission power, respectively.
For each information $d \in D_s$, the sensing cost in vehicle $s$ is denoted by $\phi_{d, s}$.
The distance between vehicle $s$ and RSU $e$ is denoted by $\operatorname{dis}_{s,e}^t$.
The sensing information indicator indicating whether information $d$ is sensed by vehicle $s$ at time $t$, is denoted by 
\begin{equation}
	c_{d, s}^t \in \{0, 1\}, \forall d \in D_{s}, \forall s \in S, \forall t \in T
\end{equation}
Thus, the set of information sensed by vehicle $s$ at time $t$ can be denoted by $D_s^t = \{ d | c_{d, s}^{t} = 1, \forall d \in D_s \}, D_s^t \subseteq D_s$, where the sensing frequency and uploading priority are denoted by $\lambda_{d,s}^t$ and $p_{d,s}^t$, respectively.
Due to the limited sensing ability, we have the following constraints on information sensing. 
\begin{equation}
	\lambda_{d,s}^{t} \in [\lambda_{d,s}^{\min} , \lambda_{d,s}^{\max} ], \ \forall d \in D_s^t, \forall s \in S, \forall t \in T
\end{equation}
\begin{equation}
	{p}_{d^*, s}^t \neq {p}_{d, s}^t, \forall d^* \in D_s^t \setminus \left\{ d\right \}, \forall d \in D_s^t, \forall s \in S, \forall t \in T
\end{equation}
where $\lambda_{d,s}^{\min}$ and $\lambda_{d,s}^{\max}$ are the minimum and maximum of sensing frequency for information with $\operatorname{type}_{d}$ in vehicle $s$, respectively.

The queuing time of information sensed by vehicles is modeled by multi-class M/G/1 priority queue \cite{moltafet2020age}.
The transmission time $\operatorname{\hat{g}}_{d, s, e}^t$ of information with $\operatorname{type}_d$ follows a class of General distribution with mean $\alpha_{d, s}^t$ and variance $\beta_{d, s}^t$.
Therefore, the uploading workload $\rho_{s}^{t}$ in vehicle $s$ is represented by $\rho_{s}^{t}=\sum_{\forall d \subseteq D_s^t} \lambda_{d,s}^{t} \alpha_{d, s}^t$.
According to the principle of the multi-class M/G/1 priority queue, it requires $\rho_{s}^{t} < 1$ to guarantee the existence of the queue steady-state.
The inter-arrival time ${i}_{d, s}^t$ is the duration between the arrival of two adjacent information with $\operatorname{type}_d$ in vehicle $s$, i.e., ${i}_{d, s}^t={1}/{\lambda_{d, s}^{t}}$
Therefore, the arrival moment and updating moment of the freshest information with $\operatorname{type}_d$ before time $t$ are denoted by $\operatorname{a}_{d,s}^t$ and $\operatorname{u}_{d,s}^t$, respectively, which can be obtained by $\operatorname{a}_{d,s}^t={ \lfloor t  \lambda_{d,s}^t  \rfloor} {i}_{d, s}$ and $\operatorname{u}_{d,s}^t = \lfloor  {\operatorname{a}_{d,s}^t}/{u_d} \rfloor u_d$, where $u_d$ is the updating interval.
The set of information with a higher uploading priority than information $d$ is denoted by $D_{d, s}^t = \{ d^* \mid p_{d^*,s}^{t} > p_{d,s}^{t} , \forall d^* \in D_s^t \}$.
Thus, the uploading workload ahead of information $d$ in vehicle $s$ at time $t$ is denoted by $\rho_{d, s}^{t}=\sum_{\forall d^* \in D_{d, s}^t} \lambda_{d^*,s}^t \alpha_{d^*, s}^t$.
According to the Pollaczek-Khintchine formula \cite{takine2001queue}, the queuing time of information $d$ in vehicle $s$ is calculated by 
\begin{equation}
    \text{\footnotesize$\operatorname{q}_{d, s}^t= \frac{1} {1 - \rho_{d,s}^{t}} 
        \left[ \alpha_{d, s}^t + \frac{ \lambda_{d,s}^{t} \beta_{d, s}^t + \sum\limits_{\forall d^* \in D_{d, s}^t} \lambda_{d^*,s}^t \beta_{d^*, s}^t }{2\left(1-\rho_{d,s}^{t} - \lambda_{d,s}^{t} \alpha_{d, s}^t\right)}\right] 
        - \alpha_{d, s}^t$}
\end{equation}

\subsection{Data Uploading Model}

We model the data uploading via reliability-guaranteed V2I communications based on the Shannon theory.
The transmission power of vehicle $s$ at time $t$ is denoted by $\pi_{s}^t$. 
The set of vehicles within the radio coverage of RSU $e$ at time $t$ is denoted by $S_e^t=\left \{s \vert \operatorname{dis}_{s,e}^t \leq r_e, \forall s \in S \right \}, S_e^t \subseteq S$.
The V2I bandwidth allocated by RSU $e$ for vehicle $s$ at time $t$ is denoted by $b_{s, e}^t$, and we have the following constraints on data uploading.
\begin{equation}
	\pi_s^t \in \left[ 0 , \pi_s \right ], \forall s \in S, \forall t \in T
\end{equation}
\begin{equation}
	b_{s, e}^t \in \left [0,b_e \right], \forall s \in S_e^{t}, \forall e \in E, \forall t \in T
\end{equation}

The signal to noise ratio (SNR) \cite{wang2019delay} of V2I communications between vehicle $s$ and RSU $e$ at time $t$ is denoted by $\operatorname{SNR}_{s,e}^{t}=\frac{1}{N_{0}} \left|h_{s,e}\right|^{2} \tau {\operatorname{dis}_{s,e}^{t}}^{-\varphi} {\pi}_s^t
$, where $N_{0}$ is the additive white Gaussian noise (AWGN); $h_{s, e}$ is the channel fading gain; $\tau$ is a constant that depends on the antennas design; $\varphi$ is the path loss exponent, and ${\pi}_s^t$ is the transmission power of vehicle $s$ at time $t$.
According to the Shannon theory, the achievable transmission rate of V2I communications between vehicle $s$ and RSU $e$ at time $t$ is denoted by $\operatorname{z}_{s,e}^t=b_{s}^{t} \log _{2}\left(1+\mathrm{SNR}_{s, e}^{t}\right)$,
where $b_{s}^{t}$ is the bandwidth allocated by RSU $e$ at time $t$.
Thus, the transmission time of information $d$ from vehicle $s$ to RSU $e$, denoted by $\operatorname{g}_{d, s, e}^t$, is computed by 
\begin{equation}
    \operatorname{g}_{d, s, e}^t=\inf _{j \in \mathbb{R}^+} \left \{ \int_{\operatorname{k}_{d, s}^t}^{\operatorname{k}_{d, s}^t + j} {\operatorname{z}_{s,e}^t} \operatorname{d}t \geq \left|d\right| \right \} -  \operatorname{k}_{d, s}^t
\end{equation}
\noindent where $\operatorname{k}_{d, s}^t$ is the moment when vehicle $s$ starts to transmit information $d$, and $ \operatorname{k}_{d, s}^t = t +\operatorname{q}_{d, s}^t$.

We assume that the channel fading $\left|h_{s,e}\right|^{2}$ follows a class of distribution with the mean $\mu_{s,e}$ and variance $\sigma_{s,e}$.
The distribution set is represented by $\tilde{p}=\{\mathbb{P}: \mathbb{E}_{\mathbb{P}}[|h_{s, e}|^{2}]=\mu_{s,e}, \mathbb{E}_{\mathbb{P}}[|h_{s, e}|^{2}-\mu_{s,e}]^{2}=\sigma_{s,e}\}$.
The transmission reliability is measured by the possibility that a successful transmission probability is beyond a reliability threshold, i.e., $\inf_{\mathbb{P} \in \tilde{p}} \operatorname{Pr}_{[\mathbb{P}]}\left(\operatorname{SNR}_{s,e}^{t} \geq \operatorname{SNR}_{s,e}^{\operatorname{tgt}}\right) \geq \delta$, where $\operatorname{SNR}_{s,e}^{\operatorname{tgt}}$ and $\delta$ are the target SNR threshold and reliability threshold, respectively.
The set of information uploaded by vehicle $s$ and received by RSU $e$ is denoted by $D_{s, e}^{t} = \bigcup_{\forall s \in S_{e}^{t}} D_{s}^{t}$.

\subsection{Age/Cost of View Formulation}
Denote the set of views in the system as $V$, and the set of information required by view $v \in V$ is denoted by $	D_{v}=\left\{d \mid y_{d,v} = 1, \forall d \in D \right\}, \forall v \in V$, where $y_{d, v}$ is a binary indicating whether information $d$ is required by view $v$.
The number of required information in view $v$ is denoted by $|D_{v}|$.
Each view may require multiple pieces of information, i.e., $|D_{v}| = \sum_{\forall d \in D}y_{d, v} \geq 1, \forall v \in V$.
The set of views required by RSU $e$ at time $t$ is denoted by $V_e^t$.
Therefore, the set of information received by RSU $e$ and required by view $v$ can be represented by $D_{v, e}^t=\bigcup_{\forall s \in S}\left(D_v \cap D_{s, e}^t\right), \forall v \in V_e^t, \forall e \in E$, 
and $| D_{v, e}^t |$ is the number of information that received by RSU $e$ and required by view $v$,  which is computed by $| D_{v, e}^t | =  \sum_{\forall s \in S} \sum_{\forall d \in D_s} c_{d, s}^t y_{d, v}$.
Then, we define the view's five characteristics of heterogeneous information fusion, including timeliness, consistency, redundancy, sensing cost, and transmission cost.

\theoremstyle{definition}
\newtheorem{Definition}{Definition}
First, heterogeneous information is time-varying, and information freshness is essential for modeling the quality of views.
The timeliness $\Theta_{v} \in \mathbb{Q}^{+}$ of view $v$ is defined as the sum of the maximum timeliness of information sensed by each vehicle, i.e., $\Theta_{v} = \sum_{\forall s\in S_{e}^{t}} \max_{\forall d \in D_v \cap D_s^t} (\operatorname{a}_{d,s}^t + \operatorname{q}_{d,s}^t + \operatorname{g}_{d, s, e}^t-\operatorname{u}_{d,s}^{t})$.
Since different types of information have their sensing frequencies and uploading priorities, keeping the versions of different kinds of information as close as possible when constructing a view is essential. 
The consistency $\Psi_{v} \in \mathbb{Q}^{+}$ of view $v$ is defined as the maximum of the difference between information updating time, i.e., $\Psi_{v}=\max_{\forall d \in D_{v, e}^{t}, \forall s \in S_{e}^{t}} {\operatorname{u}_{d,s}^t} - \min_{\forall d \in D_{v, e}^{t}, \forall s \in S_{e}^{t}} {\operatorname{u}_{d,s}^t}$
Then, we give the formal definition of age of view, synthesizing the timeliness and consistency to measure view quality.
\begin{Definition}[\textit{Age of View, AoV}]
	The age of view $\operatorname{AoV}_{v} \in (0, 1)$ is defined as a weighted average of normalized timeliness and normalized consistency of view $v$.
	\begin{equation}
	    \operatorname{AoV}_{v} = w_1  \hat{\Theta_{v}} +  w_2 \hat{\Psi_{v}}, \forall v \in V_{e}^t, \forall e \in E
	\end{equation}
\end{Definition}
\noindent where $\hat{\Theta_{v}} \in (0, 1)$ and $\hat{\Psi_{v}} \in (0, 1)$ denote the normalized timeliness and normalized consistency of view $v$, respectively, which can be obtained by rescaling the range of the timeliness and consistency of view $v$ in $(0, 1)$ via the min-max scaling.
The weighting factors for $\hat{\Theta_{v}}$ and $\hat{\Psi_{v}}$ are denoted by $w_1$ and $w_2$, respectively, and we have $w_1+w_2=1$.
These weighting factors can be tuned accordingly based on the different requirements of upper-layer applications.

Second, vehicles may sense the same information redundantly when the view requires it, which wastes the sensing and transmission resources of the vehicles.
The redundancy $\Xi_v \in \mathbb{N}$ of view $v$ is defined as the sum of redundant information in view $v$, i.e., $\Xi_v =  \sum_{\forall d \in D_v} \left | D_{d, v, e} \right| - 1$, where $D_{d, v, e}$ is the set of the information that received by RSU $e$, required by view $v$, and has the same type with information $d$, which is represented by $D_{d, v, e}=\left\{ d^* \vert \operatorname{type}_{d^*} = \operatorname{type}_{d}, \forall d^* \in D_{v, e}^t \right \}, \forall d \in D_{v, e}^t$.
On the other hand, sensing more information also brings more cost to vehicles.
The sensing cost $\Phi_{v} \in \mathbb{Q}^{+}$ of view $v$ is defined as the sum of information sensing cost of information required by view $v$, i.e., $\Phi_{v} = \sum_{\forall s \in S_{e}^{t}} \sum_{\forall d \in D_v \cap D_s^t}{\phi_{d,s}}$.
Meanwhile, information transmission requires energy consumption of vehicles, i.e., the transmission power consumption.
The transmission cost $\Omega_{v} \in \mathbb{Q}^{+}$ of view $v$ is defined as the sum of consumed transmission power during the data uploading in view $v$, i.e., $\Omega_{v} = \sum_{\forall s \in S_{e}^{t}} \sum_{\forall d \in D_v \cap D_s^t} \pi_s^t \operatorname{g}_{d, s, e}^t$.
Then, we give the formal definitions of cost of view, which synthesize the redundancy, sensing cost, and transmission cost to evaluate the cost of view.
\begin{Definition}[\textit{Cost of View, CoV}]
	The cost of view $\operatorname{CoV}_{v} \in (0, 1)$ is defined as a weighted average of normalized redundancy, normalized sensing cost, and normalized transmission cost of view $v$.
	\begin{equation}
	    \operatorname{CoV}_{v} = w_3  \hat{\Xi_{v}} +  w_4 \hat{\Phi_{v}} + w_5 \hat{\Omega_{v}}, \forall v \in V_{e}^t, \forall e \in E
	\end{equation}
\end{Definition}
\noindent where $\hat{\Xi_{v}}\in (0, 1)$, $\hat{\Phi_{v}} \in (0, 1)$, and $\hat{\Omega_{v}} \in (0, 1)$ denote the normalized redundancy, normalized sensing cost, and normalized transmission cost of view $v$, respectively.
The weighting factors for $\hat{\Xi_{v}}$, $\hat{\Phi_{v}}$, and $\hat{\Omega_{v}}$ are denoted by $w_3$, $w_4$, and $w_5$, respectively, and we have $w_3+w_4+w_5=1$.

Given a solution $( \mathbf{C}, \bf\Lambda, \mathbf{P}, \bf\Pi, \mathbf{B} )$, where $\mathbf{C}$ denotes the determined sensing information, $\bf\Lambda$ denotes the determined sensing frequencies, $\mathbf{P}$ denotes the determined uploading priorities, $\bf\Pi$ denotes the determined transmission power, and $\mathbf{B}$ denotes the determined V2I bandwidth allocation, we formulate the problem aiming at maximizing the average view quality and minimizing the average view cost, simultaneously, which is expressed as follows:
\begin{equation}
	\begin{aligned}
		\max_{\mathbf{C}, \bf\Lambda, \mathbf{P}, \bf\Pi, \mathbf{B}} &\left\{ \frac{\sum_{\forall t \in T} \sum_{\forall e \in E} \sum_{\forall v \in V_e^t} \left(1 - \operatorname{AoV}_{v}\right)}{\sum_{\forall t \in T} \sum_{\forall e \in E} |V_e^t| } \right. \\ 
		&+ \left. \frac{\sum_{\forall t \in T} \sum_{\forall e \in E} \sum_{\forall v \in V_e^t}  (1 - \operatorname{CoV}_{v} )}{\sum_{\forall t \in T} \sum_{\forall e \in E} |V_e^t| } \right\} \label{YY}\\
	\end{aligned}
\end{equation}
\begin{subequations}
	\begin{align}
		\text { s.t. }
		& (1) \sim (3), (5), (6) \notag \\
        & \sum_{\forall d \subseteq D_s^t} \lambda_{d,s}^{t} \mu_d<1,\ \forall s \in S, \forall t \in T \tag{\ref{YY}{a}}\\
        & \inf_{\mathbb{P} \in \tilde{p}} \operatorname{Pr}_{[\mathbb{P}]}\left(\operatorname{SNR}_{s,e}^{t} \geq \operatorname{SNR}_{s,e}^{\operatorname{tgt}}\right) \geq \delta, \forall s \in S, \forall t \in T \tag{\ref{YY}{b}}\\
        & {\sum_{\forall s \in S_e^{t}}b_s^t} \leq b_e, \forall t \in T \tag{\ref{YY}{c}}
	\end{align}
\end{subequations}
where (\ref{YY}{a}) guarantees the queue steady-state; (\ref{YY}{b}) guarantees transmission reliability, and (\ref{YY}{c}) requires that the sum of V2I bandwidth allocated by the RSU $e$ cannot exceed its capacity $b_e$.

\section{Proposed Solution}

\begin{figure}
\centering
  \includegraphics[width=1\columnwidth]{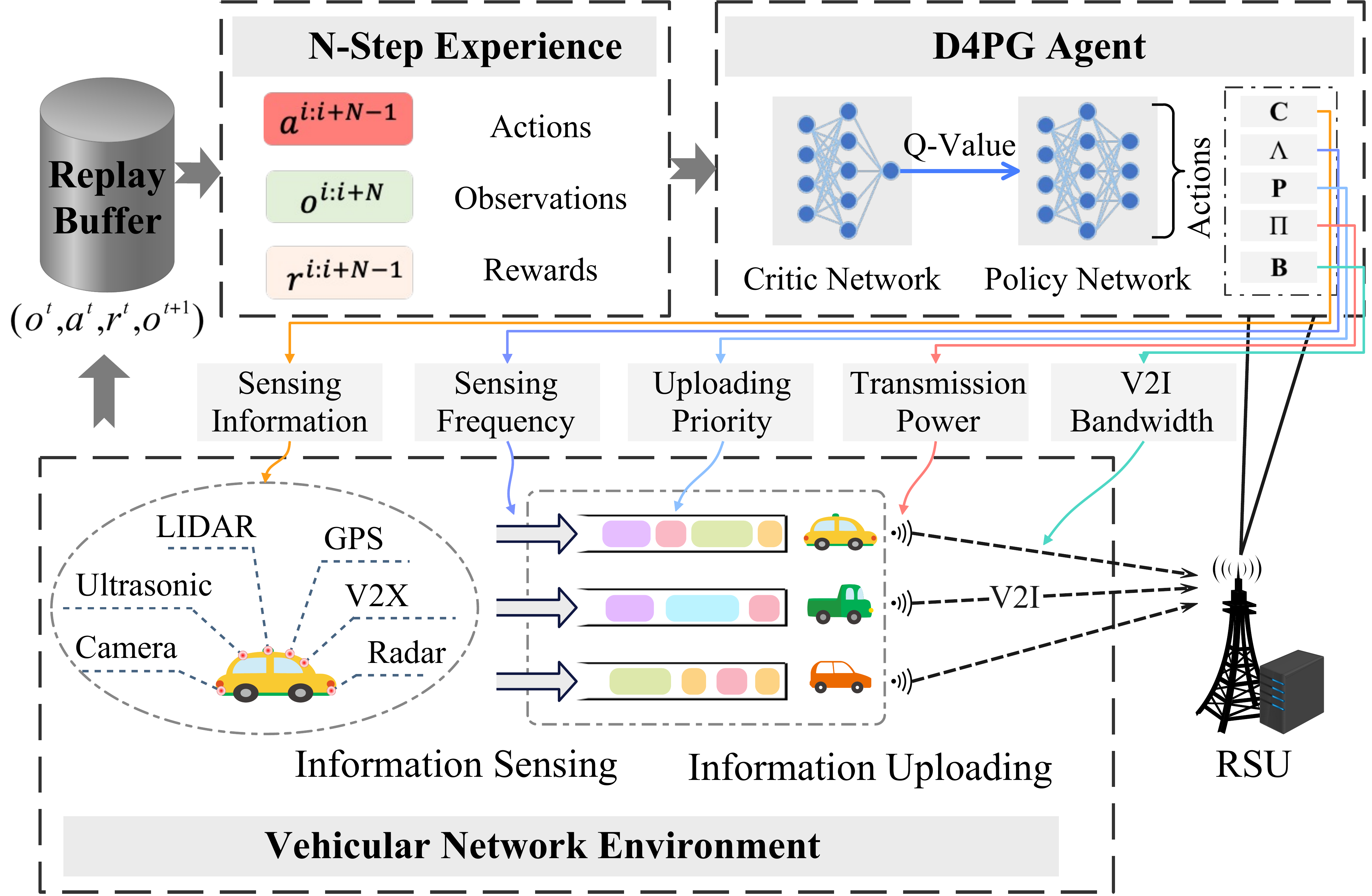}
  \caption{Distributed distributional deep deterministic policy gradient}
  \label{fig solution model}
\end{figure}

In this section, we propose the D4PG model as shown in Fig. \ref{fig solution model}, implemented in each RSU to jointly determine the sensing information, sensing frequency, uploading priority, transmission power, and V2I bandwidth.
The D4PG of RSU $e$ consists of four networks, namely, the local policy network, local critic network, target policy network, and target critic network.
The parameter of the local policy and critic networks in RSU $e$ denoted by $\theta_e^{\mu}$ and $\theta_e^{Q}$, respectively, are randomly initialized.
Then, the parameters of target policy and critic networks are initialized as the same as the corresponding local network, which are denoted by $\theta_e^{\mu^{\prime}}$ and $\theta_e^{Q^{\prime}}$, respectively.
And the replay buffer $\mathcal{B}$ is initialized to store replay experiences.

The initialized system state of each iteration is denoted by $\boldsymbol{o}^{0}$.
The local observation of the system state in the RSU $e$ at time $t$ is denoted by 
	\begin{equation}
		\boldsymbol{o}_{e}^{t}=\left\{t, e, \operatorname{Dis}_{S, e}^{t}, D_{1}, \cdots, D_{s}, \cdots, D_{|S|}, D_{e}^{t}, V_{e}^{t} \right\}
	\end{equation} 
	\noindent where $t$ is the time slot index;
	$e$ is the RSU index; 
	$\operatorname{Dis}_{S, e}^{t}$ represents the set of distances between vehicles  and RSU $e$;
	$D_{s}$ represents the set of information that can be sensed by vehicle $s$;
	$D_{e}^{t}$ represents the set of cached information in RSU $e$ at time $t$,
	and $V_e^t$ represents the set of views required by RSU $e$ at time $t$.
	Thus, the system state at time $t$ can be denoted by $\boldsymbol{o}^{t}=\left\{\boldsymbol{o}_{1}^{t}, \ldots, \boldsymbol{o}_{e}^{t}, \ldots, \boldsymbol{o}_{E}^{t}\right\}$.
The action of RSU $e$ at time $t$ is obtained based on the local observation of the system state:
\begin{equation}
	\boldsymbol{a}_{e}^{t}={\mu}\left(\boldsymbol{o}_{e}^{t} \mid \theta^{\mu}_{e}\right)+\epsilon  \mathcal{N}_{t}
\end{equation}
\noindent where $\mathcal{N}_{t}$ is an exploration noise to increase the diversity of RSU actions, and $\epsilon$ is an exploration constant.

Then, the action space of RSU $e$ consists of the offloading decision of tasks requested by vehicle $v \in \mathbf{V}_{e}^{t}$, which is denoted by 
\begin{equation}
	\boldsymbol{a}_{e}^{t} = \left \{\left \{ C_s^t, \left \{ \lambda_{d, s}^{t}, p_{d, s}^{t} \mid \forall d \in D_{s}^t \right \} , \pi_s^t   \right \}, b_{s, e}^{t} \mid \forall s \in S_{e}^{t} \right \}
\end{equation}
\noindent where $C_s^t$ is the sensing information decision; $\lambda_{d, s}^{t}$ and $p_{d, s}^{t}$ are the sensing frequency and uploading priority of information $d \in D_s^t$, respectively. 
$\pi_s^t$ is the transmission power of vehicle $s$ at time $t$, and $b_{s, e}^t$ is the V2I bandwidth allocated by RSU $e$ for vehicle $s$ at time $t$.
The set of RSU actions is denoted by $\boldsymbol{a}^{t} = \left\{\boldsymbol{a}_{e}^{t}\mid \forall e \in \mathbf{E} \right\}$. 
The actions of RSUs $\boldsymbol{a}^{t}$ are executed in vehicular network environment.
The objective of each RSU is to maximize its view quality and minimize the cost.
Therefore, the reward function of the RSU $e$ is defined as the sum of the complement of achieved average AoV and CoV of RSU $e$ at time $t$, which is represented by 
\begin{equation}
	r_{e}^{t} = {\sum_{\forall v \in V_e^t} \left(2 - \operatorname{AoV}_{v} - \operatorname{CoV}_{v}\right)} / { |V_e^t| }
		\label{edge_reward}
\end{equation}
The set of rewards of RSUs is denoted by $\boldsymbol{r}^{t} = \{r_{1}^{t}, \ldots, r_{e}^{t}, \ldots, r_{E}^{t}\}$.

Finally, the interaction experiences including the system state $\boldsymbol{o}^{t}$, RSU actions $\boldsymbol{a}^{t}$, rewards of RSUs $\boldsymbol{r}^{t}$, and next system state $\boldsymbol{o}^{t+1}$ are stored into the replay buffer $\mathcal{B}$.
A minibatch of $M$ transitions of length $N$ is sampled from replay buffer $\mathcal{B}$ to train the policy and critic networks. 
The transition of the $M$ minibatch is denoted by $\left(\boldsymbol{o}^{i:i+N}, \boldsymbol{a}^{i:i+N-1}, \boldsymbol{r}^{i:i+N-1}\right)$.
The target distribution of RSU $e$ is denoted by $Y_e^i$, which is computed by 
\begin{equation}
	Y_e^{i} = \sum_{n=0}^{N-1} \left( \gamma^{n} r_{e}^{i+n}\right)+\gamma^{N} Q^{\prime}\left(\boldsymbol{o}_{e}^{i+N}, \boldsymbol{a}_e^{i+N} \mid \theta^{Q_e^{\prime}} \right)
\end{equation}
\noindent where $\boldsymbol{a}_{e}^{i+N}$ is obtained via the target policy network, i.e., $\boldsymbol{a}_{e}^{i+N} = \mu^{\prime}(\boldsymbol{o}_{e}^{i+N} \mid \theta_e^{\mu^{\prime}})$.
The loss function of the critic network is represented by the following:
\begin{equation}
	{L}\left(\theta_e^{Q}\right)=\frac{1}{M} \sum_{i}  \left(Y_e^{i}-Q\left(\boldsymbol{o}_{e}^{i}, \boldsymbol{a}^{i} \mid \theta_e^{Q}\right)\right)^{2}
\end{equation}
The parameters of the policy network are updated via policy gradient.
\begin{equation}
	\nabla_{\theta_e^{\mu}} \mathcal{J} = \frac{1}{M} \sum_{i} \nabla_{\boldsymbol{a}_{e}^{i}} Q\left(\boldsymbol{o}_{e}^{i}, \boldsymbol{a}_e^{i} \mid \theta_e^{Q}\right) \nabla_{\theta_e^{\mu}} \mu\left(\boldsymbol{o}_{e}^{i} \mid \theta_e^{\mu}\right)
\end{equation}
The local policy and critic network parameters are updated with the learning rate $\alpha$ and $\beta$.
Finally, the RSUs update the parameters of target networks if $t \mod t_{\operatorname{tgt}} = 0$, 
\begin{equation}
	\begin{aligned}
			\theta_e^{\mu^{\prime}} &\leftarrow n \theta_e^{\mu}+(1-n)  \theta_e^{\mu^{\prime}}\\
			\theta_e^{Q^{\prime}} &\leftarrow n  \theta_e^{Q}+(1-n) \theta_e^{Q^{\prime}}
	\end{aligned}
\end{equation}
\noindent where $t_{\operatorname{tgt}}$ is the target network parameter updating period, and with $n \ll 1$.

\section{Numerical Results}

In this section, we validate the proposed solution to evaluate the performance.
In our system, we consider $E = 9$ RSUs are uniformly distributed in a 3$\times$3 $\operatorname{km}^2$ square area, where the realistic vehicular trajectories are collected from Didi GAIA open data set \cite{didi} by extracting from Qingyang District, Chengdu, China, on 16 Nov. 2016.
The information sizes are uniformly distributed in $\left|d\right| \sim$ [100 B, 1 MB], and we set the transmission power as $\pi_{s}=$ 100 mW.
We consider the additive white Gaussian noise, path loss exponent, and channel fading gain as $N_{0}=$ -90 dBm, $\varphi=$ 3, and $h_{s, e}\sim$ [2-mean, 0.4-variance] distributions \cite{wang2019delay}, and the communication bandwidth of RSU is set to $b_e=$ 20MHz.
The weighting factors for $\hat{\Theta_{v}}$ and $\hat{\Psi_{v}}$ are set as $w_1=0.6$ and $w_2=0.4$, and the weighting factors for $\hat{\Xi_{v}}$, $\hat{\Phi_{v}}$, and $\hat{\Omega_{v}}$ are set as $w_3=0.2$, $w_4=0.4$, and $w_5=0.4$.

For the implementation of the D4PG, the architectures of the policy and critic networks are described as follows.
The local policy network is a five-layer fully connected neural network with three hidden layers, where the number of neurons is 256, 256, and 256, respectively.
The architecture of the target policy network is the same as the local policy network.
The local critic network is a five-layer fully connected neural network with three hidden layers, where the numbers of neurons are 512, 512, and 256, respectively.
The architecture of the target critic network is the same as the local critic network.

For performance comparison, we implement two comparative algorithms, namely, random allocation (RA), which randomly selects one action to determine the sensing information, sensing frequencies, uploading priorities, transmission power, and V2I bandwidth allocation, and multi-agent deep deterministic policy gradient (MADDPG) \cite{zhang2021adaptive}, which is implemented in vehicles to decide the sensing information, sensing frequencies, uploading priorities, and transmission power based on local observation of the physical environment, and the RSU to determine the V2I bandwidth allocation.

\begin{figure}[htp]
\centering
  \includegraphics[width=0.7\columnwidth]{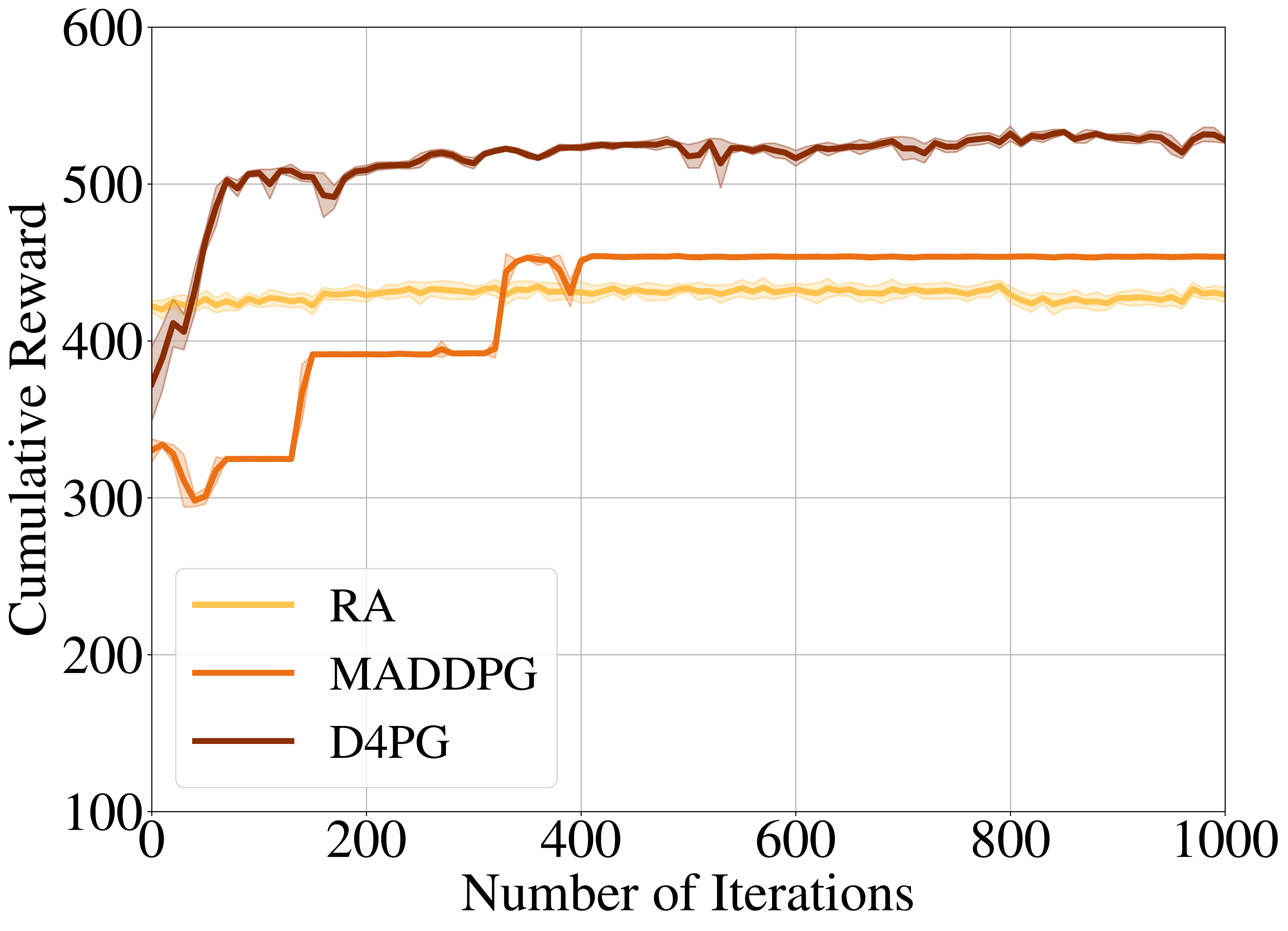}
  \caption{Convergence comparison}
  \label{fig algorithm convergence}
\end{figure}

\begin{figure}[htp]
\centering
  \includegraphics[width=0.95\columnwidth]{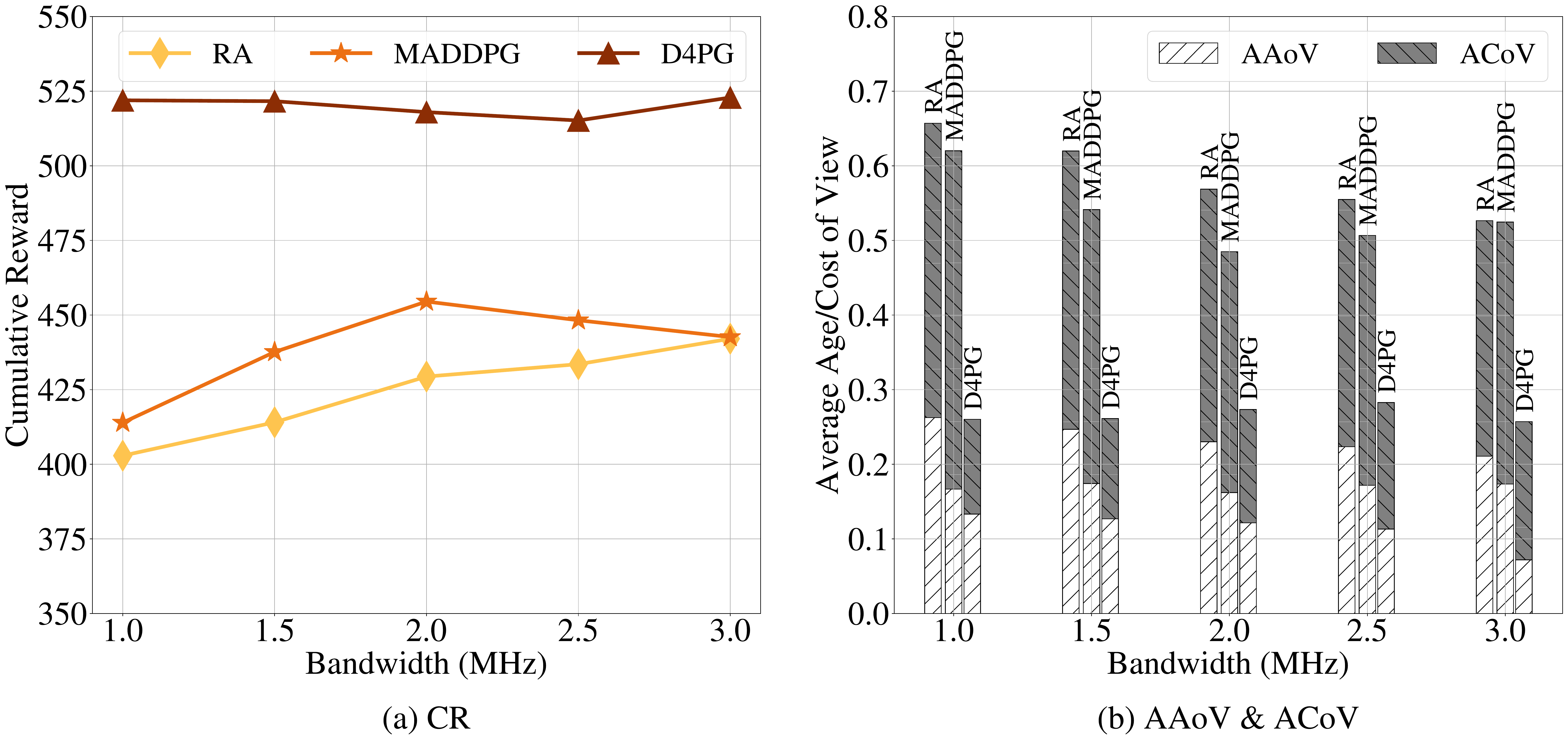}
  \caption{Performance comparison under different V2I bandwidths}
  \label{fig different bandwidth}
\end{figure}

\begin{figure}[htp]
\centering
  \includegraphics[width=0.95\columnwidth]{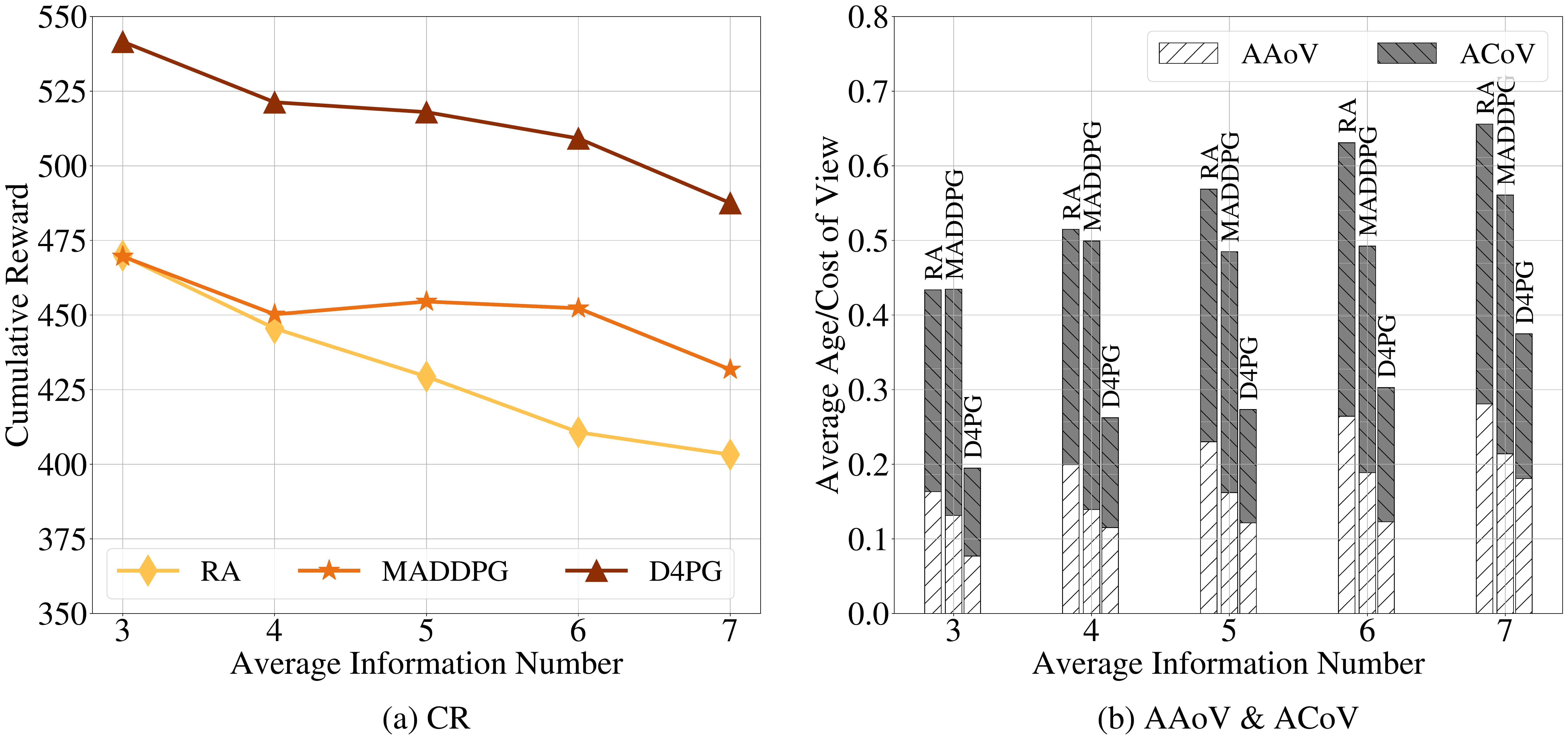}
  \caption{Performance comparison under different information numbers}
  \label{fig different number}
\end{figure}

To compare the algorithm convergence, Fig. \ref{fig algorithm convergence} compares the cumulative reward (CR) of the three algorithms.
As noted, D4PG converges the fastest (around 300 iterations) and achieves the highest CR (around 517).
In comparison, RA and MADDPG achieve a CR of around 428 and 454, respectively.
Figure \ref{fig different bandwidth} compares the performance of the three algorithms under different V2I bandwidths.
A larger bandwidth represents that the allocated bandwidth for each vehicles can be enlarged, which results in a shorter uploading time.
As the bandwidth increases, the CR of RA increases accordingly.
It is noted that the CR of MADDPG increases when the bandwidth increases from 1 MHz to 2 MHz and decreases when the bandwidth increases from 2 MHz to 3MHz.
The reason is that the system reward consists of two conflicting objectives, i.e., the AoV and CoV, which can be verified in Fig. \ref{fig different bandwidth}(b) compared the average AoV (AAoV) and average CoV (ACoV) of views.
Figs. \ref{fig different bandwidth}(a) and \ref{fig different bandwidth}(b) show that D4PG can achieve the best performance across all cases.
Figure \ref{fig different number} compares the performance of the three algorithms under different average information numbers of view requirements.
The larger average number of required information for the views indicates that the vehicles have a higher workload in information sensing and uploading, which leads to a poorer quality of views. 
With the increasing average required information number, the CR for all algorithms decreases accordingly.

\section{Conclusion}
In this paper, an information sensing model was modeled based on multi-class M/G/1 priority queue, and a data uploading model was modeled based on reliability-guaranteed V2I communications.
On this basis, two new metrics AoV and CoV were designed to evaluate the quality and cost for the logical views of VCPS.
Then, the bi-objective problem was formulated to maximize the quality and minimize the cost of VCPS modeling.
Further, the D4PG solution was proposed to jointly determine the sensing information, sensing frequencies, uploading priorities, transmission power, and V2I bandwidth allocation.
Finally, the comprehensive performance evaluation demonstrated the superiority of the proposed solution.

%

\bibliographystyle{IEEEtran} 
\bibliography{reference}

\end{document}